\documentclass[twocolumn,showpacs,amsfonts,amssymb,amsmath,floats,aps,superscriptaddress]{revtex4-1}
\usepackage{graphicx}

\usepackage{dcolumn}
\usepackage{multirow}
\usepackage{color}

\usepackage[normalem]{ulem} 

\usepackage[outdir=./]{epstopdf}
\usepackage{upgreek}

\def\expect#1{\left\langle#1\right\rangle}

\usepackage{bm}
\usepackage[colorlinks]{hyperref}
\renewcommand{\vec}[1]{\bm{#1}}

\DeclareGraphicsExtensions{.pdf,.eps}
\graphicspath{{figs/}}

\begin{document}

\title{Electron Spin Noise under the Conditions of Nuclei Induced Frequency Focusing}

\author{Natalie J\"aschke}

\address{Lehrstuhl f\"ur Theoretische Physik II, Technische Universit\"at Dortmund,
Otto-Hahn-Stra{\ss}e 4, 44227 Dortmund, Germany}

\author{Frithjof B.\ Anders}
\address{Lehrstuhl f\"ur Theoretische Physik II, Technische Universit\"at Dortmund,
Otto-Hahn-Stra{\ss}e 4, 44227 Dortmund, Germany}

\author{Mikhail M. Glazov}
\address{Ioffe Institute, 194021 St.\-Petersburg, Russia}

\date{\today}

\begin{abstract}
We study theoretically the electron spin noise in quantum dots under non-equilibrium conditions caused by the pumping by a train of circularly polarized optical pulses. In such a situation, the nuclear spins are known to adjust in such a way, that the electron spin precession frequencies become multiples of the pump pulse repetition frequency. This so called phase synchronization effect was uncovered in [Science {\bf 317}, 1896 (2007)] and termed nuclei-induced frequency focusing of electron spin coherence. Using the classical approach to the central spin model we evaluate the nuclear spin distribution function and the electron spin noise spectrum. We show that the electron spin noise spectrum consists of sharp peaks corresponding to the phase synchronization conditions and directly reveal the distribution of the nuclear spins. We discuss the effects of nuclear spin relaxation after the pumping is over and analyze the corresponding evolution of nuclear spin distributions and electron spin noise spectra.
\end{abstract}
\maketitle

\section{Introduction}

The effects of optical electron spin control in semiconductor quantum dots 
are highly topical nowadays~\cite{dyakonov_book,gywat2010spins}. The primary focus lies in
the non-magnetic initialization, manipulation and control of single spins~\cite{mikkelsen07,Press:2010aa,Arnold2015}. Due to rather weak interaction of a single spin with light, the realization of such a single spin device is rather 
challenging.
In this regard, the control of the spin dynamics in quantum dot ensembles
appears to be the much more favorable route~\cite{yakovlev_bayer,glazov:review}.
However, the inevitable inhomogeneity in the quantum dot ensembles results in 
an efficient decoherence of the electron spins due to the nuclear fluctuations and the spread of the electron Land\'e factors.

It turns out that the role of the
inhomogeneity can be effectively reduced by the
nuclei-induced electron spin precession frequency focussing effect
~\cite{Greilich28092007}. 
Under the excitation of the quantum dot ensemble by a periodic train of circularly polarized pulses with the repetition period $T_\mathrm{R}$ the nuclear spins adjust in such a way 
that the electron spin precession frequency in each dot becomes a multiple of the repetition frequency $\pi/T_\mathrm{R}$. 

The origin of the focussing is the hyperfine coupling of electron and nuclear spins~\cite{Greilich28092007,Carter2009,yakovlev_bayer,glazov:review}
accounted for by the central spin model  (CSM) \cite{Gaudin1976}.
Several quantum-mechanical \cite{petrovYakovlev,BeugelingUhrigAnders2016,BeugelingUhrigAnders2017}
or semiclassical 
\cite{Greilich28092007,Carter2009,PhysRevB.85.041303,Korenev2011,springerlink:10.1134/S1063776112060131,petrovYakovlev,JaeschkeAnders2017}
approaches have been proposed in order to describe this frequency focussing effect.

In the steady state of the periodically driven system, the Overhauser field distribution  approaches a non-equilibrium  function that significantly differs from its Gaussian shape in equilibrium \cite{petrovYakovlev,BeugelingUhrigAnders2016,BeugelingUhrigAnders2017,JaeschkeAnders2017}.
This distribution function, which is crucial for testing the 
theoretical predictions and obtaining further control over the electron spin dynamics, is difficult to  access directly in experiments.

Measurements of the electron spin dynamics \cite{Greilich28092007,JaeschkeAnders2017}  
have been performed only on the periodically driven system where the pumping affects the nuclear spin states. 
Here we suggest an alternative route to access the nuclear spin distribution by studying the electron spin fluctuations using the electron spin noise spectroscopy technique. 
Spin fluctuations can be measured by an off-resonant optical beam which does not significantly perturb the system~\cite{aleksandrov81,PhysRevLett.80.3487,Crooker_Noise,Oestreich_noise}, see Refs.~\cite{Zapasskii:13,Oestreich:rev,2016arXiv160306858S} for reviews. 
We calculate the spectrum of the electron spin fluctuations under the conditions of the nuclei-induced frequency focussing effect. We demonstrate that the electron spin noise directly reveals the distribution of the nuclear spins. We show that by monitoring the electron spin fluctuations after the train of pump pulses 
has stopped the slow nuclear spin relaxation  towards the equilibrium state can be accessed.

After a short introduction of the model in Sec.\ \ref{sec:CSM}, Sec.~\ref{sec:spin_noise} demonstrates the relation between the electron spin dynamics and the shape of the Overhauser field distribution. 
We also present the evaluation of the electron spin noise and compare it with the Overhauser field distribution. The effect  of the nuclear spin relaxation and its influence on the electron spin noise spectra
are discussed on the last part of the Sec.~\ref{sec:spin_noise}.
 A brief summary of the results is given in Sec.~\ref{sec:conclusion}.

\section{Central spin model for nuclei-induced electron spin precession frequency focussing}
\label{sec:CSM}

In this section we formulate the semiclassical model of the electron and nuclear spin dynamics in a quantum dot under periodic optical excitation.
 
In a pump-probe experiment a negatively charged semiconductor quantum dot is subjected to periodic laser pulses. We consider the Voigt geometry, i.\ e.\ the external magnetic field $\vec{B}_\mathrm{ext} \parallel x$ is orthogonal to the light propagation direction $z$ which is also the growth axis of the quantum dot. The electron spin dephasing between two 
pulses is governed by the hyperfine interaction with the nuclear spin bath, which acts as an effective magnetic field. 
The circularly polarized pump pulse excites a $X^-$ trion state consisting of two electrons in a spin singlet and a hole. Depending on the helicity of light the resident electron becomes polarized, see Refs.~\cite{PhysRevLett.96.227401,PhysRevB.80.104436,Glazov2013:rev,yakovlev_bayer,JaeschkeAnders2017} for details.

The classical equations of motion for 
the spin dynamics of the  central spin model 
subject to an instantaneous laser pulse are given by \cite{Glazov2013:rev,PhysRevB.85.041303}
\begin{subequations}
\label{eq:SCA-EOM}
\begin{eqnarray}
\label{eq:central-spin}
\dfrac{\mathrm{d}\vec{S}}{\mathrm{d}t} &=& \left( \vec{b}_\mathrm{ext} + \sum_{k=1}^N a_k 
\vec{I}_k \right) \times \vec{S} + \gamma P_{T}\mathrm{e}^{-2\gamma t}\bm e_z\\
\dfrac{\mathrm{d}\vec{I}_k}{\mathrm{d}t} &=& \left( \zeta\vec{b}_\mathrm{ext} + a_k\vec{S}\right) \times \vec{I}_k. \label{eq:nuclear-spin}
\end{eqnarray}
\end{subequations} 
These equations are valid between two consecutive laser pulses for the time $t\in [0, T_\mathrm{R})$, where $T_\mathrm{R}$ is the pump pulse repetition period. The term $\gamma P_{T}\mathrm{e}^{-2\gamma t}$ represents the increase of the electron spin vector in $z$-direction
due to the trion decay into the electron state. The quantity $P_\mathrm{T}$ is the efficiency of the trion photogeneration,
and $\gamma$ is the trion decay rate.

We treat the electron spin, $\bm S$, and the spins of the nuclei, $\bm I_k$, as classical vectors. The subscript $k$ enumerates the $N$ nuclei  interacting with the electron spin. 
The applicability of the classical approach is justified in Refs.~\cite{PhysRevB.76.045312,JaeschkeAnders2017}.

The coupled differential equations~\eqref{eq:SCA-EOM} 
are written in dimensionless units
using the characteristic time scale $T^*$ \cite{merkulov}, 
\begin{eqnarray}
\label{eq:tstar}
\frac{1}{(T^*)^2} &=& \sum_{k=1}^N A_k^2 \expect{I^2_k},
\end{eqnarray}
where we use $ \expect{I^2_k} =1$ in the classical simulation.
This time scale is determined by the fluctuations of the Overhauser field 
where $A_k$ is the hyperfine
coupling constant of the $k^\mathrm{th}$ nuclear spin $\vec{I}_k$ to the central electronic spin $\vec{S}$. 
In Eqs.~\eqref{eq:SCA-EOM}, the quantities $a_k = A_k T^*$ are the dimensionless coupling constants and the dimensionless magnetic field acting on the electron spin is $\vec{b}_\mathrm{ext} = g_\mathrm{e}\mu_\mathrm{B}\vec{B}_\mathrm{ext}T^*$, with $\mu_B$ being the Bohr magneton and $g_{\rm e}$ being the electron $g$-factor. The parameter $\zeta$ denotes
the ratio of the nuclear and electron magnetic moments. 

Equation~\eqref{eq:central-spin} describes the electron spin precession in the total field comprising the external magnetic field and the field of the nuclei as well as the electron spin generation by the pump pulses: The efficiency of the trion photogeneration $P_\mathrm{T}$ is obtained from the electron spin $z$-component before the pump pulse arrival, $S_{\mathrm{bp},z}$, as 
\begin{align}
P_{T} = S_{\mathrm{bp},z} + \frac{1}{2},
\end{align}
for the ideal $\pi$-pulses considered in this paper.
The electron spin after the pulse can by calculated via
\begin{align}\label{eq:pulse_S}
\vec{S}_\mathrm{ap} = \frac{1}{2}\left(S_{\mathrm{bp},z}-\frac{1}{2}\right)\vec{\mathrm{e}}_z.
\end{align}
The in-plane electron spin components are erased by the $\pi$-pulse~\cite{PhysRevB.80.104436}.

The dynamics of the nuclear spin $\vec{I}_k$
is influenced by the nuclear Zeeman effect and the Knight field $a_k\vec{S}$, see Eq.~\eqref{eq:nuclear-spin}. The pump pulses do not produce any direct coupling to the nuclear spins so that the nuclear spin directions before and after the pump pulse arrival remain unchanged.

For a sufficiently long train of pump pulses a steady state situation is reached: the change of the nuclear spin vectors averaged over $T_\mathrm{R}$ vanishes and the electron spin dynamics becomes periodic. 
In this limit, the dimensionless Overhauser field in Eq.\ \eqref{eq:central-spin},
\begin{eqnarray}
\label{b:N}
\vec{b}_N &=& \sum_{k=1}^N a_k \vec{I}_k 
\end{eqnarray}
is replaced by a constant vector.
The analysis of Eq.~\eqref{eq:SCA-EOM} in this frozen Overhauser field approximation 
demonstrates that the electron spin  steady state condition requires that  one
of two following conditions is satisfied \citep{JaeschkeAnders2017}: 
\begin{subequations}
\label{eq:resCond}
\begin{eqnarray}
\label{eq:even}
\omega = \omega_{\mathrm e, n}  & \equiv & \frac{2\pi n}{T_\mathrm{R}}  \\
\label{eq:odd}
\omega = \omega_{\mathrm o, n}  &\equiv & \frac{1}{T_\mathrm{R}}\left[ 2\pi n + 2 \arctan\!\left(\dfrac{\omega}{2\gamma}\right)\right],~~n\in\mathbb{Z} .
\end{eqnarray}
\end{subequations}
where $\omega$ is the electron spin precession between the pulses, 
\begin{equation}
\label{omega:def}
\omega = \left|\bm b_\mathrm{ext} + \vec{b}_N
\right|.
\end{equation} 
Note, that even in the steady state the Overhauser field varies slightly due to the nuclear spin precession, particularly in a strong magnetic field.
The rate, however, is  much slower than $\omega$ justifying the frozen Overhauser field approximation.

The condition in Eq.~\eqref{eq:even} depends only on the external magnetic field and the value of the pulse repetition rate $T_\mathrm{R}$, which corresponds to the so-called even resonances, i.\ e.\ even multiples of $\pi/T_\mathrm{R}$,
while the condition in Eq.\  \eqref{eq:odd}
also includes the influence of the trion decay. For large external magnetic fields, ${b}_\mathrm{ext}/\gamma\gg 1$,
Eq.\ \eqref{eq:odd} leads to resonance conditions with odd integer numbers $\omega_{\mathrm o, n} T_\mathrm{R} = (2n+1)\pi $. 

For the even resonance condition the central spin is fully aligned in the negative or positive $z$ direction after the pulse whereas the odd resonance conditions give a spin alignment of $\vec{S}_\mathrm{ap} = {\mp}\vec{\mathrm{e}}_z/6$ depending on the helicity of light~\cite{PhysRevB.85.041303, BeugelingUhrigAnders2016}. The detailed study of the electron and nuclear spin dynamics in the presence of periodic pumping is given in Ref.~\cite{JaeschkeAnders2017}. 

Starting from an unpolarized system where all spins are randomly oriented, first, the electron spin becomes polarized due to the pumping. The periodic pumping strongly amplifies the electron spin polarization in the quantum dots where Eq.~\eqref{eq:even} is fulfilled.
An electron spin revival can be observed~\cite{A.Greilich07212006,PhysRevB.80.104436,yakovlev_bayer}. 
Then, for a sufficiently long train of pump pulses, the hyperfine interaction leads to a rotation of nuclear spins and an eventual nuclear polarization in the direction of the external magnetic field in accordance with the resonance conditions \eqref{eq:resCond}, see ~\cite{JaeschkeAnders2017} for details. This is the nuclei-induced electron spin precession frequency focussing effect~\cite{Greilich28092007,PhysRevB.85.041303,yakovlev_bayer}. As a result, the distribution functions of the nuclear spins and, thus, the distribution of the Overhauser field

is no longer Gaussian, but becomes peaked at certain particular values of $\bm b_{\rm N}$ where the conditions~\eqref{eq:resCond} hold. The Overhauser field distribution functions $p_i(b_{\mathrm N,i})$, where $i=x$, $y$ or $z$ is the Cartesian component after the pumping stage, are shown by red solid lines in Fig.~\ref{fig:OvF_steadystate}.

\section{Spin dynamics and fluctuations}
\label{sec:spin_noise}

\begin{figure}[t]
\begin{center}
\includegraphics[width=\linewidth,clip]{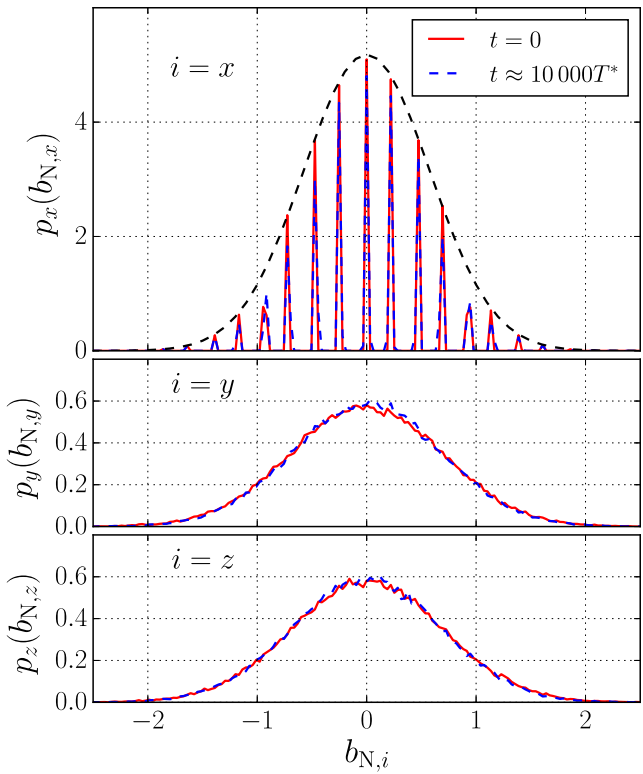}
\caption{Distributions of Cartesian components of the Overhauser field for the steady state of a periodically pulsed system. The initial distributions after the sequence of pump pulses has finished are shown in red. The Overhauser field distributions after approximately 10\,000\ $T^*$ are shown in blue. The Gaussian envelopes are depicted in black.}
\label{fig:OvF_steadystate}
\end{center}
\end{figure}

In this section we present the results of calculations of the electron spin noise after the excitation by a train of pump pulses. We also demonstrate the relation between the electron spin fluctuations and the distributions of the Overhauser field.

The numerical simulations of the electron spin noise under the conditions of the nuclei-induced frequency focussing are carried out in the following way:
We consider $N=100$ nuclear spins and simulate $N_\mathrm{C}=100\,000$ 
classical initial configurations for $\bm I_k(t=0)$ such that the distribution functions of the Overhauser field components are
given by
\begin{subequations}
\label{distrib:Overh}
\begin{align}
&p_y(b_{\mathrm{N},i}) = p_z(b_{\mathrm{N},i}) = \mathcal F(b_{\mathrm{N},i}),\\
&p_x(b_{\mathrm{N},x}) = \sum_n \mathcal F(b_{\mathrm{N},x}) \left[ \delta(b_{\mathrm{N},x}-b_{\mathrm e,n}) + \delta(b_{\mathrm{N},x}-b_{\mathrm o,n})\right],
\end{align}
\end{subequations}
where 
\begin{equation}
\label{Gauss}
\mathcal F(b_{\mathrm{N},i}) = \frac{1}{\sqrt{2\pi \sigma^2}}\exp\left(-\frac{b_{\mathrm{N},i}^2}{2\sigma^2}\right) \text{ for }i=x,y,z
\end{equation} 
is the Gaussian distribution with a variance $\sigma = 1/3$ providing the electron spin decoherence, $\delta(b)$ is the Dirac $\delta$-function and $b_{\mathrm e, n}$ [$b_{\mathrm o,n}$] is the nuclear field which satisfies the condition~\eqref{eq:even} [\eqref{eq:odd}]. 

We assume that the coupling constants are distributed according to 
\begin{align}\label{eq:distribution_Ak}
p_A(A) = -\frac{3}{2 r_0^3}\frac{1}{A}\sqrt{\mathrm{ln}\left(\frac{A_\mathrm{max}}{A}\right)},
\end{align}
with the dimensionless cut-off radius $r_0 = 1.5$ and $A_{\rm max}=1$~\cite{PhysRevB.89.045317}. 
A set  $\{ A_k\} $ is drawn from the distribution $p_A(A)$ and normalized 
via $a_k = A_k T^*$ using the definition \eqref{eq:tstar}.
Further, we chose  $T_\mathrm{R} = 13.5\,T^*$ for the repetition time of the pulses
to make contact to the experiment
and $\vec{b}_\mathrm{ext} = 2\pi K / T_\mathrm{R} \vec{\mathrm{e}}_x$ with $K = 200$ for the external field
corresponding to a physical value of approximately $2$\,T for typical quantum dot parameters. The relative strength of the nuclear Zeeman is given by $\zeta=0.00125$, and the trion decay rate is set to $\gamma = 10/T^*$.

In our simulation the restriction for the generation of the initial Overhauser field is an accepted deviation of 
\begin{equation}
\label{dev}
|\Delta b_{\mathrm{N},x}| = \left|\left(\sum_k a_k \bm I_{k,x} \right) - b_{\mathrm{o/e},n}\right|<10^{-3},
\end{equation}
 from the peak positions $b_{\mathrm{o/e},n}$ in each configuration which leads to a small but finite peak width instead of an ideal $\delta$-function peak.

We investigated the evolution of the nuclear steady state in darkness, i.\ e.\ after
stopping the pump pulses. At $t=0$, the steady state distribution
defined in Eq.\ \eqref{distrib:Overh} is assumed, and the last pump pulse arrives. 
The peak width at this time is due to restrictions of the generation method, described  in Eq.\ \eqref{dev}. 
This finite peak width mimics experiments in which a perfect frequency focusing cannot be achieved due to different nuclear spin relaxation mechanism or fluctuations in the pumping laser.
In Fig.~\ref{fig:OvF_steadystate}, we present the comparison between the initial steady-state Overhauser field 
distribution and its evolution after $t=1000 T_\mathrm{R} \approx 10\,000T^*$. 
Only marginal changes are observable, and the peak structure is essentially preserved on this time scale.

This phenomenon will be analysed more deeply in the following sections, 
with the additional focus on the electron spin dynamics.

\subsection{Electron spin dynamics after the pumping}\label{subsec:electron}

\begin{figure}[t]
\begin{center}
\includegraphics[width=0.7\linewidth,clip, angle=90]{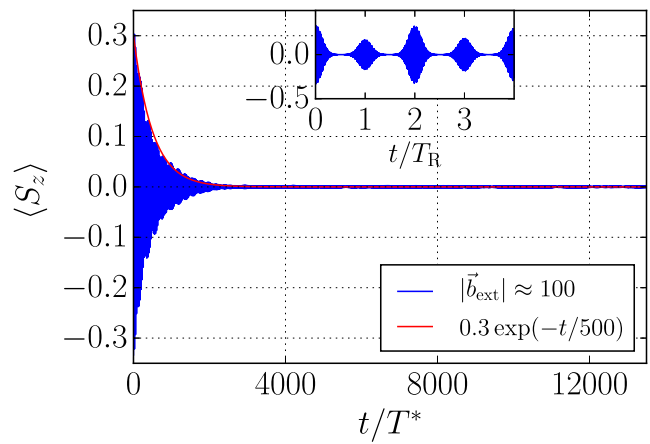}
\caption{Central spin dynamics starting from the steady state of a pulsed system. The red line gives the envelope of the electron spin revival decay. The inset shows spin dynamics on the shorter time scale.}
\label{fig:CSt_steadystate}
\end{center}
\end{figure}

Figure~\ref{fig:CSt_steadystate} shows the electron spin dynamics $\langle S_z(t)\rangle$ after the last pump pulse using the  Overhauser field distribution introduced in the previous section.
The electron spin precession in the total magnetic field $\bm b_{\rm ext}+\bm b_{\rm N}$ is very fast so that field-induced spin beats are not resolvable in the figure.
On a long time scale, the electron spin envelope function
decays with time due to a finite width of peaks in $p_x(b_{{\rm N,}x})$. 
Indeed, if the Overhauser field peaks are approximated by a Lorentzian function, the long-time decay of the electron spin polarization envelope is described by the exponential law $\langle S_z(t)\rangle \sim \exp{(-t/T_\mathrm{D})}$ with $T_\mathrm{D}$ 
being the decay time inversely proportional to the Lorentzian width. 
We added an exponential function with $T_\mathrm{D}=500T^*$  (red solid line) as a guide to the eye 
into Fig.\ \ref{fig:CSt_steadystate} demonstrating a reasonable agreement with the envelope function of our simulation.
Naturally, different peak shapes provide different time envelopes of electron spin beats. In our simulations the value of $T_\mathrm{D} \sim 10^{3} T^*$ is limited by the accuracy of the randomly generated Overhauser field distribution, Eq.~\eqref{dev}.

On a time scale of a few $T_\mathrm{R}$, the revivals of the electron spin polarization 
are depicted in the inset to Fig.~\ref{fig:CSt_steadystate} 
resolving the electron spin dynamics on a much shorter time scale  up to $t=4T_\mathrm{R}$.
These revivals at integer multiples of $T_\mathrm{R}$ 
are a clear signature of the nuclei-induced spin precession frequency focussing: 
The most probable spin precession frequencies are, in agreement with Eqs.~\eqref{eq:resCond}, the multiples of $\pi/T_\mathrm{R}$. The alternating revival strengths shown in the inset  result from the 
interplay of the even and odd resonance frequencies. 
For the larger revival amplitudes the spin configurations with even and odd resonance frequencies 
align: $\vec{S}_\mathrm{even} = \mp \vec{\mathrm{e}}_z/{2}$ and 
$\vec{S}_\mathrm{odd} =  \mp \vec{\mathrm{e}}_z/6$ (for $\sigma^+/\sigma^-$ pumping), whereas for every second $T_\mathrm{R}$ distance the contributions due to even and odd resonances point in the opposite directions: $\vec{S}_\mathrm{even} = \mp\vec{\mathrm{e}}_z/2$ and $\vec{S}_\mathrm{odd} = \pm \vec{\mathrm{e}}_z/6$.

While the central spin shows a clearly recognizable relaxation, the Overhauser field distribution is almost conserved as discussed in the previous section.
This implies that the electron spin will interact with an already prepared nuclear spin system once the pulse sequence is switched on again.
Therefore, it can be seen as an indicator for the re-emerging instant
revival after several minutes in darkness as it was observed experiments reported in Ref.~\cite{Greilich28092007}.

\subsection{Electron spin noise}

The spin noise is characterized by the second-order spin 
correlator $\langle S_z(t+t')S_z(t')\rangle$ in the time domain. 
In the frequency domain, the electron spin noise spectrum is obtain by its Fourier transformation:
\begin{align}
(S_z^2)_\omega (t') = \int_{-\infty}^\infty \mathrm{e}^{-\mathrm{i}\omega t} \langle S_z(t+t')S_z(t')\rangle  \mathrm{d}t
\, .
\end{align} 
We stress that the spin noise spectrum is dependent
on the absolute time $t'$ due to the non-equilibrium, non-stationary 
situation:
at $t'=0$ the nuclear spin system is described by the distribution functions~\eqref{distrib:Overh}, 
which is strongly different from the equilibrium 
Gaussian, and where an average electron spin polarisation is 
found - see Fig.\ \ref{fig:CSt_steadystate}.

However, electron and nuclear spin dynamics are characterized by 
time scales which differ by several orders of magnitude.
The electron spin dephasing takes place on a time scale of $T_\mathrm{D}$ related with incomplete focussing, see Sec.~\ref{subsec:electron}. In contrast, the nuclear spin relaxation characterized by the time constant $T_{1,\mathrm{N}}$ is extremely slow and can last for tens of minutes. Hereafter we assume that 
\begin{equation}
\label{condition}
T_{1,\mathrm{N}}\gg t' \gg T_\mathrm{D},
\end{equation}
 which allows us to consider the system in quasi steady state:
The spin correlation function 
$\langle S_z(t+t')S_z(t')\rangle$ and the spin noise spectrum $(S_z^2)_\omega$ 
become independent on the time $t$.
In agreement with the general theory of non-equilibrium spin fluctuations~\cite{ll10_eng,springerlink:10.1007/BF02724353} the correlation functions of the fluctuations obey the same set of equations as the average values. Thus, Eq.~\eqref{eq:central-spin} (with $P_\mathrm{T}=0$) can be considered as the equation for the electron spin correlators $\langle S_\alpha(t+t')S_z(t')\rangle$ as functions of time 
$t$. Also, it has to be taken into account that the nuclear spin dynamics in Eq.~\eqref{eq:nuclear-spin} is driven by the electron spin fluctuations. The initial nuclear spin distribution is given by~\eqref{distrib:Overh}, while the equal time electron correlation functions are given by
 $\langle S_z(t)S_z(t)\rangle=1/4$,
and the cross correlators vanish: 
$\langle S_y(t)S_z(t)\rangle= \langle S_x( t)S_z(t)\rangle=0$, which follows from the quantum-mechanical definition of the electron spin. Note, that due to condition~\eqref{condition} the initial correlation of the electron spin orientation and Overhauser field can be disregarded. The electron spin correlators and spin noise spectra can weakly depend on $t'$ due to the nuclear spin dynamics.
 
\begin{figure}[t]
\begin{center}
\includegraphics[width=\linewidth,clip]{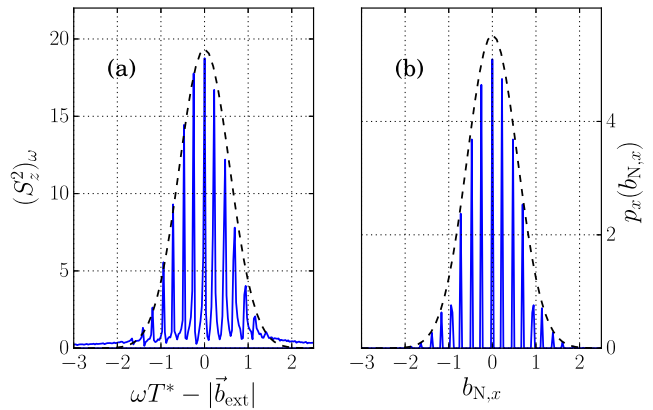}
\caption{
(a) Spin noise spectrum of the full numerical simulation.
(b) The initial Overhauser field distribution
taken from
 Fig.\ \ref{fig:OvF_steadystate} at $t'=1000T_\mathrm{R}\approx 10\,000T^*$. 
}  

\label{fig:SpinNoise_OvF}
\end{center}
\end{figure}

In order to fulfill the requirement $T_\mathrm{D}\ll t'$, we used as initial condition
the Overhauser field distribution 
obtained by a full simulation of the EOM
for $t'=1000T_\mathrm{R}\approx 10\,000T^*$ after the pump pulses have stopped --
see Fig.\ \ref{fig:OvF_steadystate}.
Figure~\ref{fig:SpinNoise_OvF}(a) shows the spin noise spectrum $(S_z^2)_\omega(t')$ 
calculated by the full numerical solution of the set of Eqs.~\eqref{eq:SCA-EOM}
and the non-equilibrium Overhauser field distribution. 
The spin noise spectrum consists of a series of peaks at the electron spin precession frequencies $\omega$  satisfying the resonance conditions Eq.~\eqref{eq:resCond}. The calculations demonstrate that the electron spin correlation function and, correspondingly, the spin noise spectrum is almost independent of 
$t$, implying a very slow nuclear spin relaxation.

The fact that the nuclear fields are almost static, 
allows for the evaluation of
the spin noise spectrum semiclassically:
The model developed in 
Ref.~\cite{Glazov2012} assumes that the nuclear fields are frozen. 
A similar model has been employed in Refs.~\cite{PhysRevB.91.205301,Ryzhov:2016aa} to address the electron spin fluctuations in the presence of dynamical nuclear polarization. In this model the electron spin noise spectrum directly reflects the distribution of the effective magnetic field acting on the electron spin. It is given, up to a common factor, by
\begin{equation}
\label{Sz2:class}
(S_z^2)_\omega(t') \propto p_x\left(\omega - b_{\mathrm{ext},x} \right).
\end{equation}
In order to show the connection between the full numerical simulation and the
analytical prediction, the
corresponding initial
distribution of the nuclear fields is added as
Fig.~\ref{fig:SpinNoise_OvF}(b) being in a good agreement with the full calculation presented in the panel~(a). Slight discrepancies are related 
to neglecting the nuclear spin dynamics in Eq.~\eqref{Sz2:class} and to
 the contributions to the electron spin precession frequency of the nuclear field components transversal to the external magnetic field. This analysis confirms that the information about the nuclear spin polarization can be accessed by the spin noise spectrum, $(S_z^2)_\omega(t')$.

\subsection{Inclusion of a phenomenological relaxation of the Overhauser field}\label{sec:phen}

\begin{figure}[t]
\begin{center}
\includegraphics[width=\linewidth,clip]{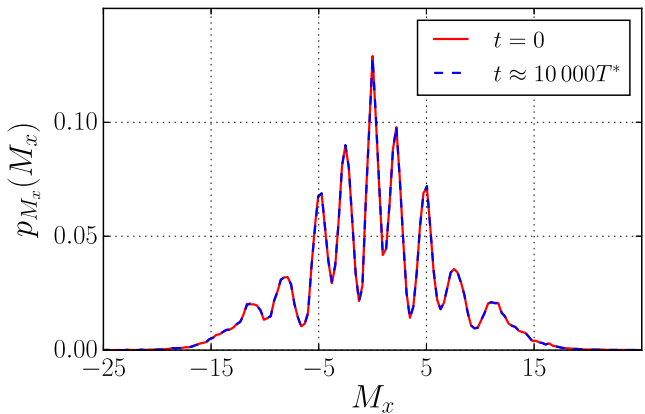}
\caption{The distribution of the total nuclear spin in $x$ direction $M_x$ corresponding to the Overhauser field shown in Fig.\ \ref{fig:OvF_steadystate} is conserved. }
\label{fig:totalNuclearSpin_SS}
\end{center}
\end{figure}

As we have seen in Fig.\ \ref{fig:OvF_steadystate} the Overhauser field distribution is almost conserved for long times after the pulse sequence. The 
distribution of the coupling constants $a_k$ can induce some decay of the Overhauser field similarly to the processes described in Refs.~\cite{merkulov,PhysRevB.70.205327,CoishLoss2004,PhysRevB.76.045312}, however, this decay is quite minor.
 This can be ascribed to the fact that the component of the total spin 
\begin{align}
  \bm F = \bm S + \sum_{k} \bm I_{k},
\end{align}
 of the interacting electron-nuclear system is conserved parallel to the external field.
Since the nuclear spin bath size is large,
the total nuclear spin $\vec{M}=\sum_k \vec{I}_k \propto \sqrt{N}$ constitutes the dominating contribution to $\vec{F}$
\begin{align}
\langle F_x \rangle \approx \langle M_x \rangle = \mathrm{const}.
\end{align}   
Hence, a variation of the electron spin $x$-component can be neglected in comparison 
to the constant contribution $\langle M_x \rangle\propto \sqrt{N}$. 
It is noteworthy to distinguish the 
approximatively conserved total nuclear spin polarisation $\langle M_x \rangle$
in an external field applied in $x$-direction
from the slowly varying Overhauser field for a finite spread of the coupling constants $a_k$.
Only in the case of the box model with $a_k = a \forall k$ the difference between Overhauser field and total nuclear spin reduces to the constant $a$ and, therefore, the Overhauser field distribution is also static.
To illustrate the difference we present the 
the total nuclear spin distribution in $x$ direction, $p_{M_x}(M_x)$, in 
Fig.\ \ref{fig:totalNuclearSpin_SS} for the system parameters used in 
Fig.\ \ref{fig:OvF_steadystate}.
While a close inspection of
Fig.\ \ref{fig:OvF_steadystate} and Fig.\ \ref{fig:SpinNoise_OvF}(b)
reveals the slow time evolution of $p_x(b_{N,x})$, $p_{M_x}(M_x)$
remains constant for the simulation times $t<10\,000T^*$.

In a real system, however,
several different processes such as 
the dipole-dipole interaction between the nuclear spins~\cite{abragam},
fluctuating quadrupolar splittings of nuclei~\cite{PhysRevB.77.245201},
due to the recharging processes and the photoexcitation~\cite{Greilich28092007,PhysRevB.95.241408}
contribute to the nuclear spin relaxation that occurs on a much larger time scale as considered up to now. In order to analyse this effect 
we employ the phenomenological approach where we (i) use the box model for simplicity, (ii) calculate the electron spin noise spectrum via the
semiclassical Eq.~\eqref{Sz2:class}, and (iii) introduce the nuclear spin relaxation by means of a kinetic equation for the distribution function $p_{M_x}$.

The longitudinal nuclear spin relaxation time $T_{1,\mathrm{N}}$ is much larger than any time scale related to the electron spin dynamics. 
As a result we can regard the Overhauser field as static in a time interval while we investigate the electron spin dynamics of the correlation function $\langle S_z(t' +t)S_z(t') \rangle$. As we can gather from Fig.\ \ref{fig:CSt_steadystate} the electron spin is decayed after a relatively short amount of time. Therefore we will assume an unpolarized central spin at the beginning of each time interval starting at $t'$
with a static Overhauser field distribution at the time $t'$.

To achieve a decay of the distribution of the total nuclear spin 
component $M_x$, spin flips decreasing or increasing 
$M_x$, have to be taken into account. The general time evolution of 
the distribution function $p_{M_x}(M_x;t')$  is given by the following rate equation
\begin{align}
\label{eq:discRateEq}
\begin{split}
\dfrac{\mathrm{d}p_{M_x}(M_x;t')}{\mathrm{d}t'} &= W_\downarrow (M_x+1)p_{M_x}(M_x+1;t')\\
&+W_\uparrow (M_x-1)p_{M_x}(M_x-1;t')\\
&-\left[W_\downarrow (M_x)+W_\uparrow (M_x)\right]p_{M_x}(M_x;t').
\end{split}
\end{align}
$W_{\uparrow /  \downarrow}$ are the nuclear spin-flip probabilities, where $\uparrow$ denotes the process of increasing $M_x$ by $1$ and $\downarrow$ denotes the process of decreasing $M_x$ by $1$. Note that dipole-dipole interaction between the nuclei and the quadrupole splittings can result in the change of $M_x$ by $2$, however, the inclusion of such additional processes does not qualitatively affect the results. 

Now we employ 
the box model: $p_x(b_{\mathrm{N},x})$ and $p_{M_x}(M_x)$ differ
only by a constant prefactor. At
$t'=0$, the time at the end of the pumping, the analytic
Overhauser field distribution stated in Eq.~\eqref{distrib:Overh}
is used, and the distribution function $p_{M_x}(M_x;t')$ is normalized according to 
\begin{align*}
\sum_{M_x} p_{M_x}(M_x;t')=1
\end{align*}
at any time.
Reminding ourselves that $1\ll |M_x| \ll N$ holds in real systems,
we can convert the discrete rate equation \eqref{eq:discRateEq}
into a partial differential equation, 
\begin{align}\label{eq:ResultRateEq}
\begin{split}
&\dfrac{\partial p_{M_x}(M_x;t')}{\partial t'}= \\
&\dfrac{1}{T_{1,\mathrm{N}}}\dfrac{\partial}{\partial M_x}\left[M_x p_{M_x}(M_x;t')
 + \dfrac{M^2}{3}\dfrac{\partial}{\partial M_x}p_{M_x}(M_x;t') \right],
\end{split}
\end{align}
in the continuum limit
which is an analogue of the Fokker-Planck equation in kinetic theory~\cite{ll10_eng}. 
We made use of the expressions
\begin{align}
W_\uparrow(M_x) &= W_1 (N/2-M_x),\\
W_\downarrow(M_x) &= W_1 (N/2+M_x),
\end{align}
with $W_1$ being the probability of a single spin flip, and the factors $N/2\pm M_x$ describing the number of choices for a nuclear spin to flip, which are 
valid in the high temperature approximation~\cite{abragam}.
In Eq.\ \eqref{eq:ResultRateEq},
$T_{1,\mathrm{N}}$ 
denotes the longitudinal relaxation time of the nuclear spin governing 
the exponential decay of the average nuclear spin $\overline{M}_x(t') = \int dM_x M_x p_{M_x}(M_x;t')$:
\begin{align}
\dfrac{\partial \overline{M}_x}{\partial t'}  = -\frac{\overline{M}_x(t')}{T_{1N}},
\end{align}
and $M^2=|I^2|N$ is the square of the total spin of the nuclei. 
The steady-state solution of Eq.~\eqref{eq:ResultRateEq} corresponds to the unpolarized bath with the Gaussian distribution function of the nuclear spins,
\begin{align}
p_{M_x}(M_x) = \frac{1}{2}\sqrt{\frac{6}{\pi M^2}}\exp\left(-\frac{3M_x^2}{2M^2}\right),
\end{align}
with  the variance 
\begin{eqnarray}
\int p_{M_x}(M_x)M_x^2\mathrm{d}M_x = M^2/3.
\end{eqnarray}
For the investigation of the time dependency of
the Overhauser field distribution we use Eq.\ \eqref{distrib:Overh}  
as the initial condition.
The broadening is determined by the variance $\sigma_\mathrm{P}^2 = 10^{-4}$ 
introduced when replacing the $\delta$-functions in Eq.~\eqref{distrib:Overh} by the Gaussian as
\begin{align*}
\delta(x) \to \frac{1}{\sqrt{2\pi \sigma_{\rm P}^2}}\exp\left(-\frac{x^2}{2\sigma_{\rm P}^2}\right).
\end{align*}
As $\lim\limits_{\omega \to \infty}2\arctan [\omega/(2\gamma)] \rightarrow \pi$ in Eq.\ \eqref{eq:odd} holds for large external magnetic fields of several Tesla used in the experiments~\cite{Greilich28092007},
we set the second set of peaks exactly at odd multiples of $\pi/T_\mathrm{R}$.

\begin{figure}[t]
\centering
\includegraphics[width=\linewidth,clip]{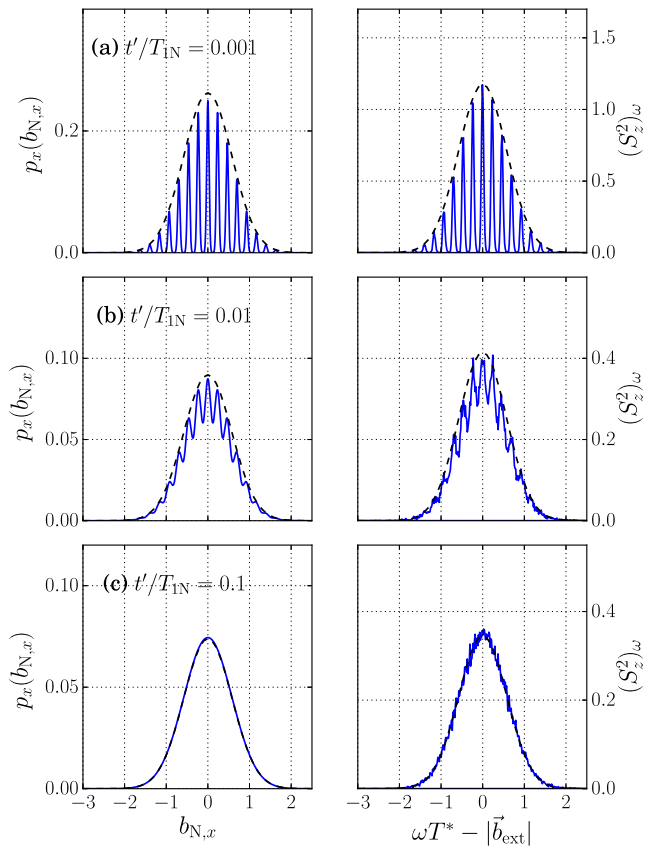} 
\caption{
Left side: normalized Overhauser field distribution. Right side
normalized spin noise spectra for (a) $t'/T_{1,\mathrm{N}}=0.001$, (b) 
 $0.01$ and (c) $0.1$.  
The spin noise spectra are shifted by the frequency of the external magnetic field for better comparability. 
}
\label{fig:OvF_SN}
\end{figure}

The differential equation \eqref{eq:ResultRateEq} can be analytically
solved by a Fourier transformation and by using a separation of variables. 
Note, that the solution of Eq.~\eqref{eq:ResultRateEq}  depends only
on the ratio $t'/T_{1,\mathrm{N}}$.  The shape of the Overhauser field distribution is calculated for the times $t'/T_{1,\mathrm{N}}=10^{-3}$, $10^{-2}$ and $10^{-1}$  
and shown in Fig.~\ref{fig:OvF_SN}, left column.

The Overhauser fields are randomly generated from these distributions for each classical configuration. 
For these different times the autocorrelation functions of the central spin $\langle S_z(t'+t) S_z(t') \rangle$ are calculated by full numerical simulations of \eqref{eq:SCA-EOM}
and their  
Fourier transformations give the corresponding
electron spin noise spectra. The results
are shown in the right column of Fig.\ \ref{fig:OvF_SN}.

The $x$ component of the Overhauser field distribution shows a clear decay of the peaked substructure inside its Gaussian envelope
driven by the nuclear spin relaxation processes parametrized by $T_{1,\rm N}$.
The electron spin noise spectrum traces remarkably the time evolution of the
Overhauser field distribution and, therefore, gives direct access to the time constant
of the nuclear spin relaxation. 
Both distributions match very well apart from normalization and the shift of the spin noise spectra by the Larmor precession frequency which is given by the external magnetic field $|\vec{b}_\mathrm{ext}|$. For longer times the nuclear spin bath relaxes to the Gaussian equilibrium state, Eq.\ \eqref{Gauss}. 
Thus, by monitoring the spin noise as a function of time $t'$ after the preparation of the nuclear spin system, one can obtain the time evolution of the nuclear spin system and extract the parameter $T_{1,\rm N}$, the nuclear spin-lattice relaxation time.

Noteworthy, a similar protocol of electron spin noise measurements can be applied in the course of the focussing process, i.e., starting from the unpolarized nuclear state and interrupting the pulse train to measure the electron spin noise. In such a case, the formation of the peaked distribution of nuclear states can be monitored and the focussing time can be estimated and compared with model predictions~\cite{Greilich28092007,PhysRevB.85.041303,Korenev2011,BeugelingUhrigAnders2016,BeugelingUhrigAnders2017,JaeschkeAnders2017}.

\section{Conclusion}
\label{sec:conclusion}

In this work we demonstrate that the distribution of the nuclear spins can be readily determined from the electron spin noise spectrum under the conditions of the nuclei-induced electron spin precession frequency focussing effect.

Based on simulations of the nuclei-spin dynamics under the effect of a pump pulse train we find the distribution of Overhauser fields acting on the electron. The distribution of the longitudinal, i.e., parallel to an external magnetic field, component of the Overhauser field features sharp peaks at the resonant conditions, where the electron spin precession frequency in the total field (being sum of external and Overhauser field), is commensurable with the pump pulse repetition frequency, weighted by a Gaussian envelope.

It was shown that the Overhauser field distribution
 is stable for a macroscopically long time after the end of pulsing
 due to the angular momentum component conservation in the electron-nuclear spin system. The calculated electron spin noise spectrum closely follows the Overhauser
distribution function.

We also study the evolution of the electron spin noise spectrum driven by the slow nuclear spin relaxation, which is accounted for by a simple kinetic equation. We demonstrate that the nuclear relaxation toward the structureless Gaussian distribution is also directly revealed in the electron spin fluctuations.


\begin{acknowledgments}

We are grateful to A.\ Fischer, I.\ Kleinjohann and P.\ Schering for discussions. We acknowledge the financial support by the
Deutsche Forschungsgemeinschaft and the Russian Foundation of Basic
Research through the transregio TRR 160. M.M.G. is grateful to RFBR projects 17-02-00383 and 15-52- 12012 and Russian President grant MD-1555.2017.2 for partial support. The authors gratefully acknowledge the computing
time granted by the John von Neumann Institute for Computing (NIC) under project HDO09 and provided on the supercomputer JUQUEEN  at the J\"ulich Supercomputing Centre.
\end{acknowledgments}

\appendix

%

\end{document}